# Human-centric Maintenance Process Through Integration of AI, Speech, and AR

Parul Khanna
Department of Civil, Environmental and Natural Resources Engineering
Luleå University of Technology
0920-492386
parul.khanna@ltu.se

Ravdeep Kour
Department of Civil, Environmental and Natural Resources Engineering
Luleå University of Technology
0920-492344
ravdeep.kour@ltu.se

Ramin Karim
Department of Civil, Environmental and Natural Resources Engineering
Luleå University of Technology
0920-492898
ramin.karim@ltu.se

## ABSTRACT
The adoption of Augmented Reality (AR) is increasing to enhance Human-System Interaction (HSI) by creating immersive experiences that improve efficiency and safety in various industries. In industrial maintenance, traditional practices involve physical documentation and device interactions, which might disrupt the task, affect efficiency, and increase the cognitive load for the maintenance personnel. AR has the potential to support and enhance industrial maintenance processes in these aspects. Therefore, the purpose of this research is to study and explore how advanced technologies like Artificial Intelligence (AI), AR and speech processing can be integrated to support hands-free, real-time task logging and interaction in maintenance environments. This is done by developing a demonstrator for Microsoft HoloLens 2 using Unity, C#, Azure Cognitive Services, and Azure Functions, which enables speech-to-text transcription for hands-free maintenance support. Using Azure's speech recognition, the demonstrator can achieve high transcription accuracy in an AR environment, facilitating natural interactions between users and the augmented environment. The study aims to explore the potential of AR to reduce cognitive load, streamline workflows, and improve safety by enhancing HSI for maintenance personnel in high-stakes environments.

## 1. INTRODUCTION
As industries embrace the digital transformation, integrating Augmented Reality (AR) and Human-System Interaction (HSI) is redefining the way maintenance operations are conducted, enhancing their efficiency, accuracy and safety. HSI plays a critical role in successful industrial operations, as it focuses on how humans interact with increasingly complex systems. The main aim is to create seamless, intuitive and user-friendly industrial workflows that minimise errors and enhance performance [1]. AR has emerged as a powerful tool within the HSI domain, capable of overlaying digital information onto the physical environment to create immersive, interactive experiences [2], [3]. By integrating virtual elements with the real world, AR enables enhanced informed decision-making and operational efficiency [4].

There is potential for using AR in industrial contexts, especially in domains like training, assembly, manufacturing, and maintenance. AR helps reduce the use of hardcopy manuals or handheld devices by overlaying visual instructions on the user's field of view in the real world. This reduces task interruptions and enhances safety [5]. In the maintenance domain, AR facilitate technicians to access the maintenance logs and procedural guidance digitally. This eliminates the need to carry physical things during maintenance tasks, which enhances task efficiency, safety and reduces the cognitive load of the technicians [6].

Maintenance itself is a structured process involving several key stages, including inspection, fault detection, diagnosis, corrective or preventive actions, and documentation [7]. Although the industries are embracing advanced technologies, most of the maintenance processes still rely on traditional manual documentation and post-task reporting, which is usually carried out at the end of the maintenance personnel's shift. Such an approach can introduce risks of inaccuracy, especially in fault reporting, repair records, and the sequence of events. It can cause inconsistencies in fault occurrences and add noise to maintenance records and can directly impact the quality of failure data analysis and data-driven decision-making. Interruptions caused by referencing hard copy manuals or using handheld devices also increase the cognitive load, which is the mental effort required to process information while performing tasks [8]. This may result in disrupted workflows, which can be a critical issue in time-sensitive or safety-critical environments [8]. Addressing these challenges requires thinking of how humans interact with tools and systems in the maintenance workflows. This highlights the importance of effective HSI to ensure accuracy, efficiency, and safety in industrial operations.

To overcome these challenges, this paper presents a developed proof-of-concept maintenance support tool that integrates AR with Artificial Intelligence (AI) and speech processing technologies. It is built utilising Microsoft HoloLens 2, Unity, C#, Azure Cognitive Services, and Azure Functions. This tool enables maintenance personnel to perform real-time, hands-free logging of tasks through natural speech interactions within an augmented environment. This approach reduces cognitive load and discrepancies caused by delayed or manual logging. This can improve data accuracy and enhance the effectiveness of fault diagnosis and predictive maintenance. This approach enhances HSI and overall safety by streamlining workflows without interrupting the continuity of the tasks.

## 2. BACKGROUND

HSI plays a critical role in balancing human capabilities with system demands, particularly in high-risk industrial domains [9]. NASA's decades of human experience during space flights demonstrate how HSI principles optimise human and technology collaboration in high-risk environments, ensuring safety and operational efficiency [10]. Research in the context of extra-vehicular space activities has demonstrated that AR interfaces can reduce attentional loss and improve procedural guidance in complex tasks [11]. Additionally, research also highlights HSI's role in minimising total ownership costs of assets while maximising system effectiveness through human-centred design [12].

The use of AR in industrial processes enhances their effectiveness and efficiency by imposing contextual data over the real world. Research has reported that AR enhances the key performance indicators such as efficiency (20–30% faster task completion), less error (15–25% reduced errors), and safety through real-time guidance [13]. However, challenges in the context of limitations and the need for further research on hardware fragmentation and user-AR interaction were also highlighted [13]. Research has also highlighted AR's compatibility with maintenance management systems to obtain the repair history in real time in a 3D environment [14]. Additionally, other studies have also demonstrated the use of AR in extreme environments, with enhanced accuracy of tasks and situational awareness in maintaining space suits [11].

Manually logging the maintenance records and inconsistencies in the reports remains a critical issue. In predictive maintenance (PdM) systems, inconsistent logging practices cause data quality issues [15]. Traditional logging methods lack adaptability to dynamic operational environments and create gaps between human performance and system demands [16]. Researchers argue that mobile AR instructions produced fewer assembly errors compared to the traditional way of following paper-based documentation [17]. The study also highlighted bad interface design leading to increased mental load and unsatisfactory user experience [17]. In turbine blade maintenance scenarios, AR assistance was found to improve maintenance task completion time by 21% and reduce mental workload by 26%, with participants preferring AR for its spatial guidance and intuitive usability [18]. Research demonstrated that the use of AR headsets while displaying maintenance instructions can enhance workers' adherence to the procedure and reduce errors compared to traditional paper-based manuals [19]. However, the study also highlights usability challenges, indicating that while AR holds promise, further refinement is needed for widespread adoption in preventive maintenance contexts. [19]

## 3. RESEARCH METHODOLOGY

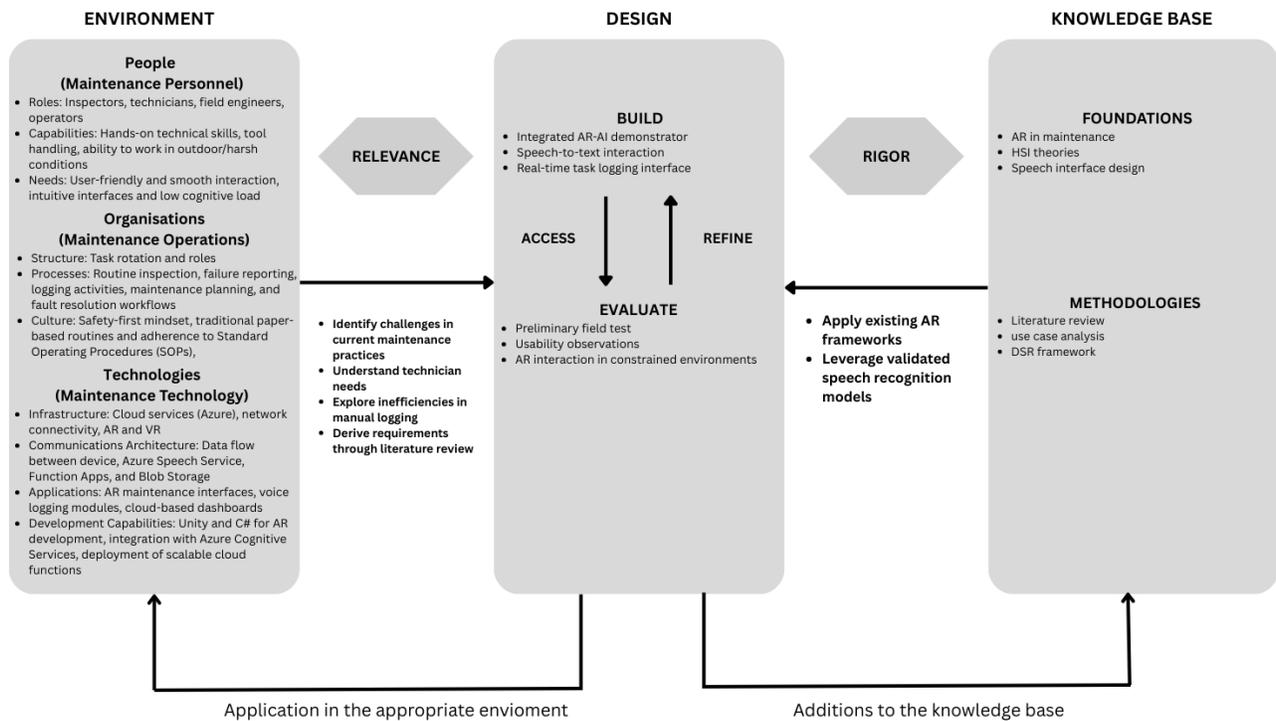

**Figure 1. Adapted DSR framework for AR-AI based Industrial Maintenance**

The conceptual framework followed in this study is based on the Design Science Research (DSR) framework proposed by [20], adapted to fit the context of industrial maintenance and the development of an integrated AR-AI solution by systematically linking real-world challenges with theory-driven design, iterative evaluation, and practical implementation (Figure 1). The framework was tailored to reflect the domain-specific characteristics of maintenance environments, involving people, organisations, and enabling technologies. In this study, the Environment comprises of three key domains:

People (Maintenance Personnel): This includes inspectors, technicians, and field engineers, whose work demands hands-on skills and the ability to operate in outdoor or challenging conditions. Their needs include smooth, intuitive interaction and minimal cognitive load during task execution.

Organisations (Maintenance Operations): These refer to task structures, shift patterns, fault logging workflows, and an organisational culture centred around safety, adherence to standard operating procedures (SOPs), and legacy documentation methods.

Technologies (Maintenance Technology): These encompass cloud-based infrastructure, in this case, Azure, AR devices (HoloLens 2), speech-to-text interfaces, and real-time communication flows between AR devices and cloud storage (Azure Functions and Blob Storage). Development capabilities include the use of Unity, C#, and AI-driven services like Azure Cognitive Services.

The Relevance of the study comes from identifying the challenges in traditional maintenance practices, such as manual documentation, task interruptions due to simultaneous fault reporting, or irregularities in fault reporting time. We used relevant keywords including "augmented reality", "maintenance", "AR in maintenance", "industrial maintenance" and their combinations and searched academic databases including Google Scholar, Scopus and ACM digital library to gather these insights. Additionally, a backward citation tracking approach was conducted to explore other relevant papers. This overall review process led to the selection of 20 relevant articles.

The Design phase focused on building the proof-of-concept demonstrator. The demonstrator enables hands-free and real-time transcription, allowing maintenance personnel to interact with the system without disengaging from the ongoing physical tasks. The Evaluate phase included a preliminary field test on a railway maintenance test rig, where the demonstrator was used in a constrained environment. For the rigorness of the DSR model, the system was developed using validated AR frameworks and speech recognition models. These ensured that both the design and evaluation were based on proven methodologies. The Knowledge Base supported the entire process as the study was conducted on established theories in AR, HSI, and speech interface design. It also encompassed the methodological choices, such as literature review, use case analysis and the application of the DSR framework itself.

Building on these insights, we propose an integrated AR-AI approach to address the observed challenges.

## 3.1 Demonstrator Architecture and Implementation

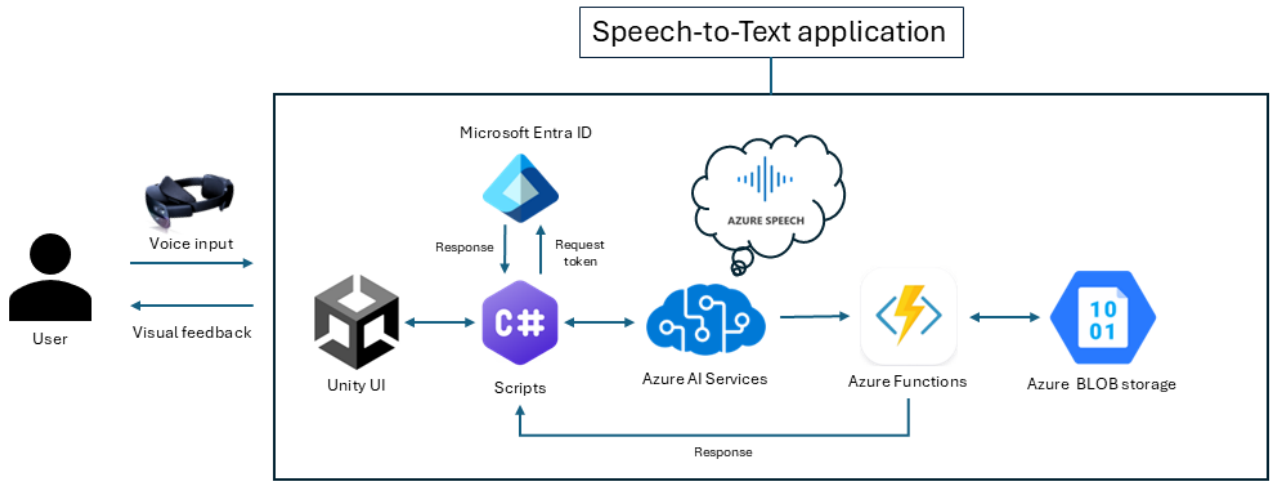

**Figure 2. System architecture of the AR-based maintenance demonstrator**

The key components of the demonstrator include the Microsoft HoloLens 2, Unity, C# and a suite of Azure cloud services. An overall flow can be seen in Figure 2. The app interaction with the user is done through the HoloLens 2, where the AR interface is built using Unity game engine and scripted in C# to manage interaction logic and cloud communication. To enable secure access and cloud integration, Microsoft Entra ID (formerly known as Azure Active Directory) was used to issue access tokens that authorise the app to use Azure services. Once authenticated, the spoken maintenance inspection logs and reports are transcribed in real time using Azure Speech Services, part of Azure's AI capabilities. The transcribed data is then passed to Azure Functions, which processes the input and stores it as a structured JSON object in Azure Blob Storage. This enables easy access, traceability, and integration with other systems.

## 4. RESULTS

The developed proof-of-concept AR demonstrator integrates Microsoft HoloLens 2 with Unity, C#, and Azure services to facilitate hands-free, real-time maintenance logging. This AR tool is designed to enhance HSI aspects and enables maintenance personnel to log maintenance inspection reports directly through speech commands. Using Azure Speech Services, spoken input is transcribed in real time, at the same time providing immediate visual feedback within the AR interface to confirm successful logging. This helps eliminate the need for physical documentation or handheld devices, reducing interruptions during task execution. Once transcribed, the data is routed through Azure Functions, where it is processed and stored as structured JSON entries in Azure Blob Storage. These records can later be accessed for reporting, tracking, or integration with other asset management systems. By linking each transcribed data with relevant metadata such as timestamps, the system supports improved traceability and fault diagnosis. This system supports seamless interaction and workflow continuity, and enhances efficiency and data reliability in industrial maintenance operations.

## 4.1 Preliminary Field Test on Railway Maintenance Rig

To demonstrate the feasibility and relevance of the developed AR-based maintenance support tool demonstrator, a small-scale preliminary field test was conducted on a railway track maintenance test rig. The purpose was to assess the system's basic functionality, responsiveness, and usability under realistic, but controlled, industrial and environmental conditions. The objective of the field test was to simulate a common maintenance task, damage inspection and logging.

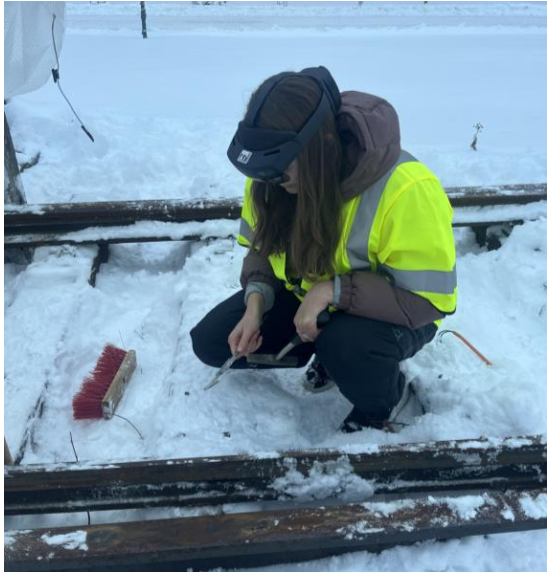

**Figure 3. Testing the demonstrator during a preliminary field test on a railway maintenance rig**

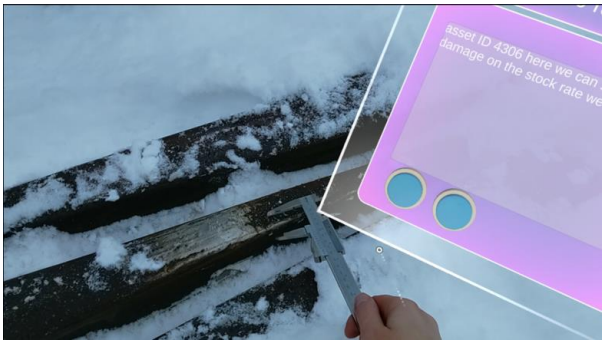

**Figure 4. Real-time transcription displayed in the AR interface during the maintenance inspection**

During the test, a team member used the AR demonstrator through Microsoft HoloLens 2 to inspect the defects on a rail track segment. As shown in Figure 3, both hands of the user are occupied with a physical inspection task to assess the damage, and the user uses the speech service to log their inspection comments and findings. The transcription was displayed in real time within the AR environment, confirming successful speech capture and system responsiveness as shown in Figure 4. This hands-free interaction validated the system's ability to maintain workflow continuity and reduce cognitive load by eliminating the need for manual data entry. Each log entry in the JSON file includes both the spoken comments and a timestamp of logging, supporting accurate and verifiable maintenance documentation. The test also allowed observation of the system's usability in extreme conditions, cold weather in this case. The practicality of speech interaction and logging over touchscreen, handheld devices and physical documentation became evident. Despite the outdoor setting and environmental constraints, the transcription service operated reliably, and no major latency or failure was observed in the cloud communication and data storage process.

The test validated that the demonstrator can support natural interaction and logging, supporting the goal of enabling uninterrupted, hands-free workflows in industrial maintenance.

## 5. CONCLUSION

This paper studied and explored the use of advanced technologies like AR, AI, and speech services to enhance HSI in the maintenance process. The developed AR-based proof-of-concept demonstrator aimed at enhancing HSI in an industrial maintenance context. The developed tool enables real-time, hands-free maintenance task logging. This allows the maintenance personnel to interact naturally with an augmented environment while keeping their hands free for other physical tasks. It helps reduce cognitive load, eliminate delays in documentation, and improve the accuracy of maintenance records. This work addressed common challenges in traditional maintenance practices, such as delayed documentation, workflow interruptions, and increased cognitive load. A preliminary field demonstration on a railway test rig showed the system's potential for natural interaction and practical usability in real-world conditions.

However, while the demonstrator shows promise, it is still in its early stages. Further work is needed to test it more extensively, refine speech recognition in noisy environments, and integrate it with existing maintenance systems. These areas present key opportunities for future work, which will focus on scaling and implementing the approach for operational deployment. Future development of the tool will include the possibility to associate the maintenance logs with specific assets. One approach can be to use embedded QR codes attached to the asset. When scanned through an AR device, the tool can retrieve the contextual asset information, such as asset ID, type, service history, or technical documentation. This asset information will then be integrated with the maintenance personnel's comments. This would enhance traceability, enable targeted fault tracking, and streamline data entry by associating asset specific metadata in the maintenance logs.

## 6. ACKNOWLEDGEMENTS

The authors gratefully acknowledge the European Commission for supporting the Marie Sklodowska Curie program through the H2020 ETN MOIRA project (GA 955681). The authors would also like to thank Emmy Erlandsson, Projektassistent at Luleå University of Technology, for her help in the initial development and testing of the demonstrator tool. Appreciation is further extended to the eMaintenance Lab at Luleå University of Technology for providing technical support and access to testing infrastructure during this work.